# Reconstruction of instantaneous flow fields from transient velocity snapshots using physics-informed neural networks: Applications to pulsatile blood flow behind a stenosis

Kakeru UEDA*, Hiro WAKIMURA* and Satoshi II*
*Department of Mechanical Engineering, School of Engineering, Institute of Science Tokyo
2-12-1 Ookayama, Meguro-ku, Tokyo 152-8550, Japan
E-mail: ii.s.148c@m.isct.ac.jp

**Abstract**
Physics-informed neural networks (PINNs) offer a promising framework by embedding partial differential equations (PDEs) into the loss function together with measurement data, making them well-suited for inverse problems. However, standard PINNs face challenges with time-dependent PDEs due to the high computational cost of space-time training and the risk of convergence to local minima. These limitations are particularly pronounced in hemodynamic analysis, where 4D-flow magnetic resonance imaging (4D-flow MRI) yields temporally sparse velocity snapshots over the cardiac cycle. To address this challenge, we propose a PINN framework that reconstructs instantaneous flow fields from transient velocity snapshots by inferring the acceleration term in the incompressible Navier-Stokes equations. By designing the network without explicit time as an input, the proposed approach enables physics enforcement using spatial evaluations alone, improving training efficiency while maintaining physical consistency with transient flow characteristics. In addition, we introduce an acceleration-mismatch loss that penalizes discrepancies between predicted and measured accelerations, which improves prediction accuracy through regularization. Numerical examples on pulsatile flow behind a stenosis using temporally and spatially downsampled synthetic data generated from time-resolved CFD demonstrate that the proposed framework reliably reconstructs velocity fields even under sparse temporal sampling, and appropriate regularization for acceleration improves predictions of pressure-gradient and acceleration fields.

## 1. Introduction

Hemodynamic assessment can benefit from high-fidelity computational fluid dynamics (CFD) simulations to obtain physiologically meaningful blood-flow features, such as spatiotemporal velocity patterns and pressure gradients. For example, the CFD velocity fields can be used to estimate hemodynamic forces acting on the vessel wall (wall shear stress), which may help predict the growth and rupture of cerebral aneurysms (Qian et al., 2011). Alternatively, imaging modalities such as four-dimensional flow magnetic resonance imaging (4D-flow MRI) (Markl et al., 2012) can noninvasively provide *in vivo* velocity data over a cardiac cycle and have become a promising source of patient-specific data for computational analyses. In practice, however, 4D-flow MRI suffers from limited spatiotemporal resolution and measurement noise, and the resulting velocity fields can be physically inconsistent. These limitations make it difficult to directly analyze the measurement data and to extract physiologically meaningful inputs (e.g., boundary conditions) for CFD simulations.

A widely used approach to address these limitations is data assimilation that combines measurement data with CFD. In particular, four-dimensional variational data assimilation (4D-Var) has shown strong capability in reconstructing *in vivo* blood-flow dynamics (Funke et al., 2018; Ichimura et al., 2025; Otani et al., 2022; Töger et al., 2020) by assimilating 4D-flow MRI velocity data into CFD simulations through optimization. Despite its accuracy, 4D-Var typically requires substantial computational resources due to the repeated forward analysis required by its adjoint formulation, which can hinder its practical use in clinical settings.

Physics-informed neural networks (PINNs) have emerged as an alternative framework for inverse problems since their introduction (Raissi et al., 2019). In standard PINNs, the neural network maps spatiotemporal coordinates to flow variables, and the governing equations are imposed through residuals computed by automatic differentiation (Baydin et al., 2018) and incorporated into the training loss together with measurement data. This formulation is attractive because it avoids explicit meshing and adjoint derivations while naturally integrating sparse measurement data, often at a lower computational cost than 4D-Var. However, for time-dependent PDEs such as the incompressible Navier-Stokes equations, standard PINNs trained over the space-time domain can still be computationally demanding, as introducing the time dimension substantially increases the number of collocation points, and training may converge slowly. These challenges may be further amplified in hemodynamic applications, where only limited velocity data is available over time.

To improve training robustness (i.e., stable and reliable convergence), a variety of training and modeling strategies have been proposed, including adaptive loss weighting (Gao et al., 2025; McClenny et al., 2023), adaptive sampling (Wu et al., 2023), adaptive activation functions (Zhang et al., 2025), and domain decompositions such as XPINNs (Jagtap and Karniadakis, 2020). Approaches tailored to time-dependent problems can further improve training stability and convergence by partitioning the time domain into a few time-slabs and training sequentially (or with interface constraints) across these slabs (Krishnapriyan et al., 2021; Mattey and Ghosh, 2022; Wight and Zhao, 2020). However, such strategies typically introduce additional algorithmic and implementation complexity (e.g., scheduling, interface coupling, and hyperparameter tuning) and may suffer from error accumulation or fail to escape local minima due to poor propagation of temporal information across time-slab interfaces (Penwarden et al., 2023).

In this study, we propose a PINN framework that targets reconstruction of an instantaneous flow field from temporally sparse velocity snapshots. The key idea is to treat the acceleration term in the incompressible Navier-Stokes equations as an unknown output variable of the network, thereby eliminating the explicit time derivative from the evaluation of the Navier-Stokes residual. Specifically, the network takes only the spatial coordinates as input and predicts the velocity, pressure, and acceleration fields at a target time instant. We also incorporate acceleration data through an acceleration-mismatch loss to regularize the network outputs.

We demonstrate the proposed framework on pulsatile flow behind a stenosis using synthetic data derived from time-resolved CFD. We examine the effects of temporal downsampling and the sensitivity to the choice of loss weights. The results demonstrate that the proposed approach is robust to temporal downsampling and to variations in loss weights when reconstructing transient velocity fields, and they also highlight the role of acceleration regularization in improving pressure-gradient and acceleration predictions. The proposed framework enables efficient reconstruction of instantaneous velocity and pressure-gradient fields at a target phase even when only temporally sparse velocity snapshots are available.

## 2. Methods

### 2.1 Pulsatile blood flow model

We model pulsatile blood flow in a vessel using the incompressible Navier-Stokes equations in non-dimensional form, consisting of the mass and momentum conservation laws:

$$\nabla \cdot \mathbf{u} = 0, \qquad \mathbf{x} \in \Omega,\ t \in [0, T], \tag{1}$$

$$\frac{\partial \mathbf{u}}{\partial t} + (\mathbf{u} \cdot \nabla)\mathbf{u} = -\nabla p + \frac{1}{Re}\nabla^2 \mathbf{u}, \qquad \mathbf{x} \in \Omega,\ t \in [0, T]. \tag{2}$$

Here, $\Omega \subset \mathrm{R}^d$ denotes the spatial domain with dimension $d = 2$ and 3, $\mathbf{x}$ is the spatial coordinate, and $t$ is time. In this study, we restrict our attention to a two-dimensional domain and set $d = 2$. The variables $\mathbf{u}(\mathbf{x}, t) \in \mathbb{R}^2$ and $p(\mathbf{x}, t) \in \mathbb{R}$ represent the non-dimensional velocity and pressure fields, respectively. $Re$ denotes the Reynolds number, defined as $Re = \rho UD / \mu$, where ρ is the fluid density, $U$ is a characteristic velocity, $D$ is a characteristic length, and $\mu$ is the dynamic viscosity. $T$ denotes the period of pulsatile flow.

The governing equations are supplemented with no-slip boundary conditions on the rigid vessel walls:

$$\mathbf{u} = \mathbf{0} \qquad \mathbf{x} \in \Gamma_{wall}, \ t \in [0, T] \tag{3}$$

where $\Gamma_{wall} \subset \partial\Omega$ denotes the vessel wall boundary.

## 2.2 Standard PINNs formulation

We first describe the standard formulation of PINNs for the incompressible Navier-Stokes equations. Figure 1 illustrates a fully connected deep neural network that maps the spatiotemporal coordinates $[\mathbf{x}, t]^T$ to the velocity and pressure field

$$\mathcal{N}_\theta : [\mathbf{x}, t]^T \mapsto [\hat{\mathbf{u}}(\mathbf{x}, t; \theta),\ \hat{p}(\mathbf{x}, t; \theta)]^T, \tag{4}$$

where $\theta$ denotes the set of trainable network parameters. The mapping $\mathcal{N}_\theta$ consists of $N$ hidden layers with $V$ neurons per layer and activation function $\sigma(\cdot)$.

The loss function consists of three main components: data loss, boundary condition loss, and physics residual loss. Given velocity data, the data loss is defined as

$$\mathcal{L}_{Data}^{vel}(\theta) = \frac{1}{N_{Data}} \sum_{i=1}^{N_{Data}} \left\| \mathbf{u}_{Data}^{i} - \hat{\mathbf{u}}(\mathbf{x}_i, t_i; \theta) \right\|^2, \tag{5}$$

where $N_{Data}$ is the number of data points, and $\mathbf{u}_{Data}^{i}$ denotes the measured velocity at the $i$-th point. To enforce the no-slip boundary condition, the boundary loss is defined as

$$\mathcal{L}_{BC}^{vel}(\theta) = \frac{1}{N_{BC}} \sum_{i=1}^{N_{BC}} \left\| \hat{\mathbf{u}}(\mathbf{x}_i, t_i; \theta) \right\|^2, \tag{6}$$

where $N_{BC}$ is the number of boundary condition points. The incompressible Navier-Stokes equations are enforced through the residuals

$$\mathcal{R}_1(\mathbf{x}, t; \theta) = \nabla \cdot \hat{\mathbf{u}}, \tag{7}$$

$$\mathcal{R}_2(\mathbf{x}, t; \theta) = \frac{\partial \hat{\mathbf{u}}}{\partial t} + (\hat{\mathbf{u}} \cdot \nabla)\hat{\mathbf{u}} + \nabla \hat{p} - \frac{1}{Re} \nabla^2 \hat{\mathbf{u}}. \tag{8}$$

The physics-based loss is then given by

$$\mathcal{L}_{PDE}(\theta) = \frac{1}{N_{PDE}} \sum_{i=1}^{N_{PDE}} \left( \| \mathcal{R}_1(\mathbf{x}_i, t_i; \theta) \|^2 + \| \mathcal{R}_2(\mathbf{x}_i, t_i; \theta) \|^2 \right), \tag{9}$$

where $N_{PDE}$ denotes the number of collocation points in the spatiotemporal domain. The total loss minimized during training is defined as

$$\mathcal{L}(\theta) = w_{Data}^{vel} \mathcal{L}_{Data}^{vel}(\theta) + w_{BC}^{vel} \mathcal{L}_{BC}^{vel}(\theta) + w_{PDE} \mathcal{L}_{PDE}(\theta). \tag{10}$$

where $w_{Data}^{vel}$, $w_{BC}^{vel}$, and $w_{PDE}$ are weighting parameters.

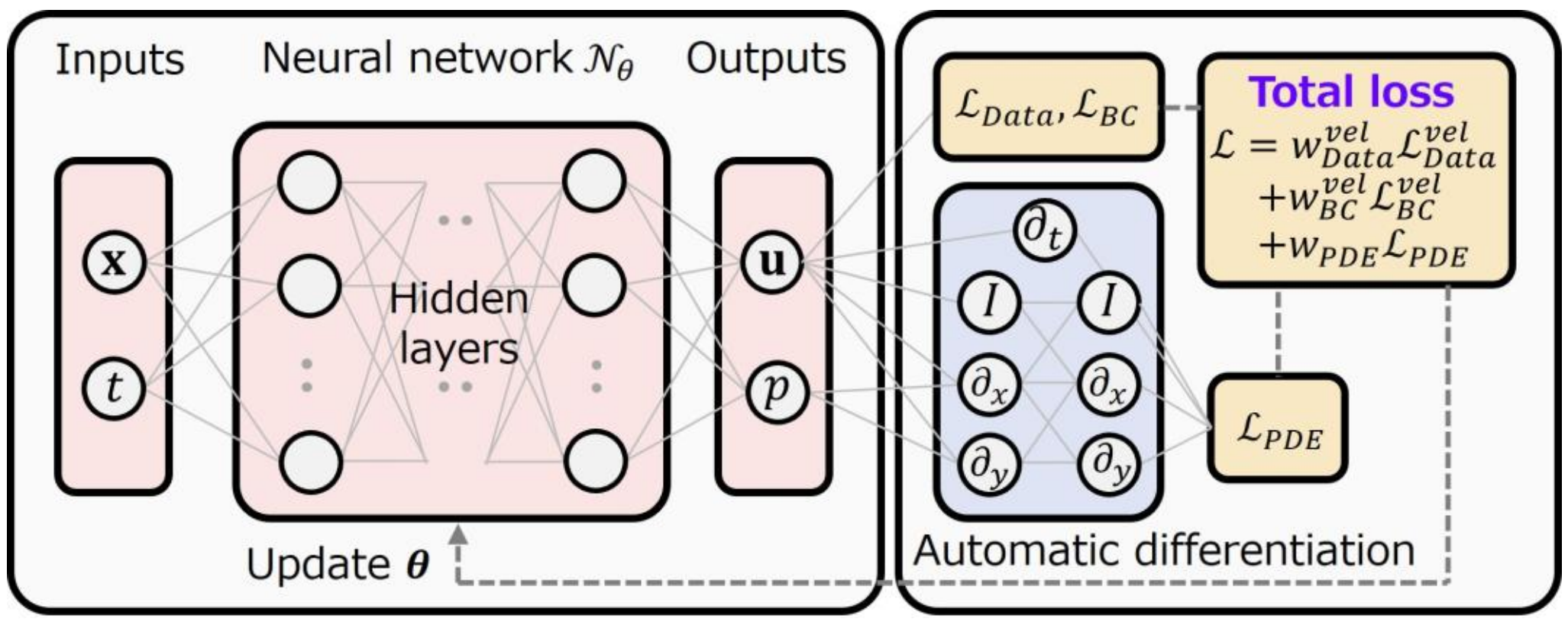

Fig. 1 Schematic diagram of the standard PINN workflow for solving inverse problems with the incompressible Navier-Stokes equations.

### 2.3 Proposed PINN framework

We propose an alternative PINN formulation for scenarios in which transient velocity snapshots are available at discrete time instants, as is typical in 4D-flow MRI applications. Figure 2 illustrates the proposed framework. Let $N_t$ velocity snapshots be available at times $\{t_k\}_{k=0}^{N_t-1}$, sampled at a uniform interval $\Delta t_{Data}$. Rather than training over the full space-time domain, we train the network on the space domain at a target snapshot time $t = t_k$ (Fig. 2(A)). The proposed network (Fig. 2(B)) takes the spatial coordinate $\mathbf{x}$ as input and predicts the velocity, pressure, and acceleration fields:

$$\mathcal{N}_{\boldsymbol{\theta}}^* : \mathbf{x} \mapsto [\hat{\mathbf{u}}(\mathbf{x};\boldsymbol{\theta}),\, \hat{p}(\mathbf{x};\boldsymbol{\theta}),\, \hat{\mathbf{a}}(\mathbf{x};\boldsymbol{\theta})]^T. \tag{11}$$

where $\hat{\mathbf{a}}(\mathbf{x};\theta)$ denotes the predicted acceleration field. Using the predicted acceleration, Eq. (8) is reformulated as

$$\mathcal{R}_2^*(\mathbf{x};\theta) = \hat{\mathbf{a}} + (\hat{\mathbf{u}} \cdot \nabla)\hat{\mathbf{u}} + \nabla\hat{p} - \frac{1}{Re}\nabla^2\hat{\mathbf{u}}. \tag{12}$$

This formulation eliminates time from the network inputs, allowing the governing equations to be evaluated using spatial derivatives alone. To provide additional guidance on the temporal variation of the velocity field, we introduce an acceleration-mismatch loss:

$$\mathcal{L}_{Data}^{acc}(\theta) = \frac{1}{N_{Data}}\sum_{i=1}^{N_{Data}} \left\|\mathbf{a}_{Data}^i - \hat{\mathbf{a}}(\mathbf{x}_i;\theta)\right\|^2. \tag{13}$$

Here, $\mathbf{a}_{Data}^i$ denotes the acceleration data at the $i$-th data point. The acceleration data is computed from the available velocity snapshots through temporal interpolation, as described in Section 2.4. Thus, the total loss for the proposed framework is defined as

$$\mathcal{L}^*(\theta) = w_{Data}^{vel}\mathcal{L}_{Data}^{vel}(\theta) + w_{Data}^{acc}\mathcal{L}_{Data}^{acc}(\theta) + w_{BC}^{vel}\mathcal{L}_{BC}^{vel}(\theta) + w_{BC}^{acc}\mathcal{L}_{BC}^{acc}(\theta) + w_{PDE}\mathcal{L}_{PDE}^*(\theta), \tag{14}$$

where $w_{Data}^{acc}$ and $w_{BC}^{acc}$ are weighting parameters, $\mathcal{L}_{BC}^{acc}$ denotes the residual enforcing no-slip boundary condition for acceleration, and $\mathcal{L}_{PDE}^*$ is constructed from the modified residual in Eq. (12).

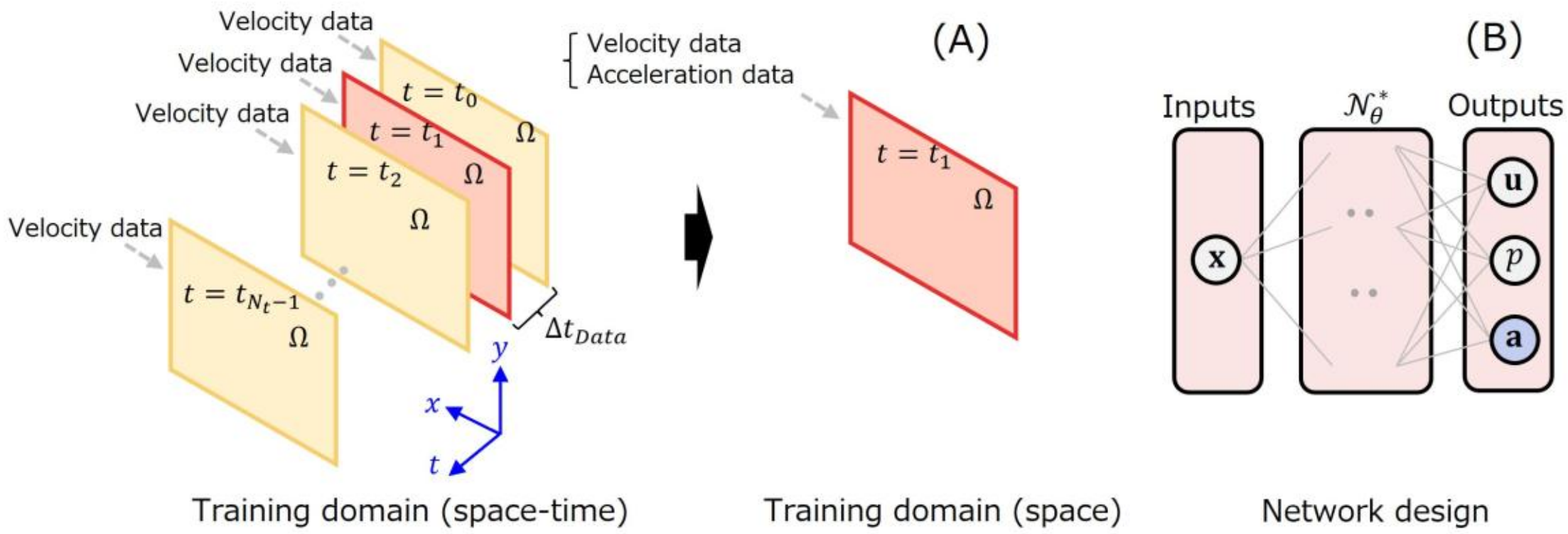

Fig. 2 Schematic diagram of the proposed PINN framework for reconstructing instantaneous flow fields from transient velocity snapshots: (A) training domain and (B) proposed network design.

### 2.4 Acceleration data

In the proposed framework, acceleration data is required to evaluate the acceleration-mismatch loss $\mathcal{L}_{Data}^{acc}$. We obtain these accelerations from the available velocity snapshots by cubic-spline interpolation in time. For a spatial sample point $\mathbf{x}_p$, we construct a cubic spline $\mathbf{s}(t) \in \mathbb{R}^2$ satisfying

$$\mathbf{s}(t_k) = \mathbf{u}_{Data}(\mathbf{x}_p, t_k), \quad k = 0, \dots, N_t - 1, \tag{15}$$

The spline $\mathbf{s}_p(t)$ is piecewise cubic and globally $C^2$ in time. To uniquely determine the cubic spline, we adopt the natural boundary condition:

$$\mathbf{s}''(t_0) \quad = \mathbf{0}, \qquad \mathbf{s}''(t_{N_t-1}) = \mathbf{0}. \tag{16}$$

The acceleration at the target time $t = t^*$ is then obtained by differentiating the spline:

$$\mathbf{a}_{Data}(\mathbf{x}_p, t^*) = \left.\frac{\mathrm{d}\mathbf{s}(t)}{\mathrm{d}t}\right|_{t=t^*}. \tag{17}$$

The accuracy of the resulting acceleration mainly depends on the temporal sampling interval $\Delta t_{Data}$.

### 2.5 Synthetic training data

Synthetic training data is generated using time-resolved computational fluid dynamics (CFD) simulations of pulsatile flow downstream of a stenosis to evaluate the proposed framework against CFD reference solutions. Figure 3 shows the computational domain Ω, an idealized expanding tube representing a simplified vessel segment with asymmetric stenosis. The domain is composed of two subdomains, $\Omega = \Omega_1 \cup \Omega_2$, where $\Omega_1 = \{(x, y) \in \mathbb{R}^2 \mid 0\ \text{mm} \leq x \leq 10\ \text{mm},\ 5\ \text{mm} \leq y \leq 10\ \text{mm}\}$ and $\Omega_2 = \{(x, y) \in \mathbb{R}^2 \mid 10\ \text{mm} \leq x \leq 100\ \text{mm},\ 0\ \text{mm} \leq y \leq 10\ \text{mm}\}$. The boundary is divided into the inlet $\Gamma_{in}$, outlet $\Gamma_{out}$, and vessel walls $\Gamma_{wall}$ (upper and lower walls).

At the inlet $\Gamma_{in}$, we impose a Dirichlet boundary condition with a time-dependent pulsatile inflow velocity. A parabolic profile is defined as

$$U(r, t) = U_{max}(t)\left(1 - \left(\frac{r}{r_0}\right)^2\right), \tag{18}$$

where $r_0$ is the vessel radius at the stenosis, and $r$ is the radial distance from the inlet center point at $(x, y) = (0, 7.5)$. $U_{max}(t)$ is the maximum value of the velocity profile defined as

$$U_{max}(t) = 0.4 + 0.2 \sin\left(\frac{2\pi t}{T}\right). \tag{19}$$

At the outlet $\Gamma_{out}$, a traction-free condition is applied, while a no-slip condition is enforced on $\Gamma_{wall}$ as described in Eq. (3).

We solve the incompressible Navier-Stokes equations using the finite element method on an unstructured triangular mesh. The equations are discretized using a Galerkin/least-squares (GLS) formulation (Thompson and Pinsky, 1995) to
enhance numerical stability. The Courant-Friedrichs-Lewy (CFL) number is kept below 1 throughout the simulations. The physical parameters are set to $\rho = 1050$ kg/m$^3$ and $\mu = 0.035$ Pa·s, which are commonly used in blood-flow simulations (Ii et al., 2018; Fillingham et al., 2023). We set the pulsation period to $T = 1$ s. This results in a peak inlet Reynolds number of $Re_{peak} = 90$ ($U = 0.6$ m/s, $L = 5$ mm) and a corresponding Womersley number of 1.1, defined as $L\sqrt{2\pi\rho/(\mu T)}$. The CFD results also serve as the reference for validating the PINN predictions and assessing reconstruction accuracy.

Figure 4(A) illustrates an example of temporally downsampled velocity snapshots obtained from CFD. For example, in one setting we retain 24 snapshots, corresponding to $\Delta t_{Data} = 0.04$ s. The velocity fields are then spatially downsampled to mimic measurement data (Fig. 4(B)). For training, we use the velocity field at $t = 0.2T$ together with the corresponding acceleration field (Fig. 4(C)). The acceleration fields are computed from the temporally downsampled velocity snapshots by cubic-spline interpolation in time at $t = 0.2T$, following Eqs. (15)-(17).

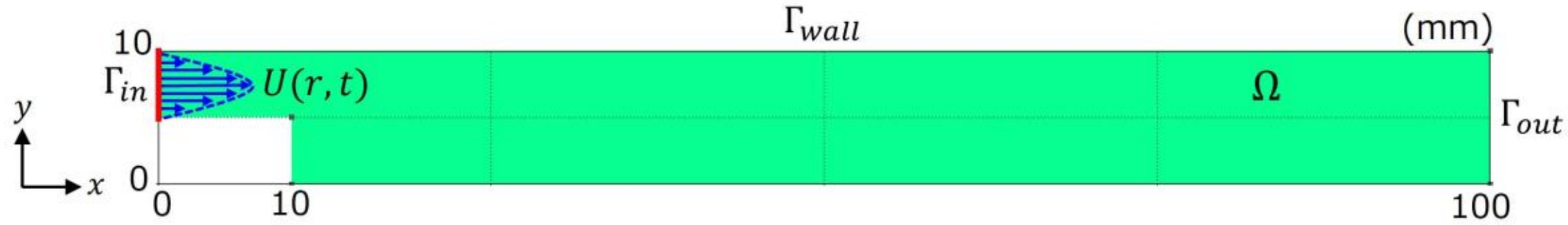

Fig. 3 Computational domain for pulsatile flow behind a stenosis.

**2.6 Training setup**

We implement the proposed PINN framework using DeepXDE (Lu et al., 2021), a Python library for training PINNs. The network is optimized with the Adam optimizer (Kingma and Ba, 2015) to minimize the total loss, using a learning rate of $10^{-4}$ for a maximum of $2 \times 10^5$ epochs. We set the network parameters to $N = 4$ hidden layers and $V = 32$ neurons per layer, and use the hyperbolic tangent activation function $\sigma(\cdot) = \tanh(\cdot)$. The network parameters are initialized with Glorot uniform initialization (Glorot and Bengio, 2010).

Figure 5 shows the spatial distribution of the training points, including boundary-condition points (500), data points (760) used for both velocity and acceleration, and PDE collocation points (5000). The physical parameters are set to match the CFD simulation: $\rho = 1050$ kg/m$^3$ and $\mu = 0.035$ Pa·s.

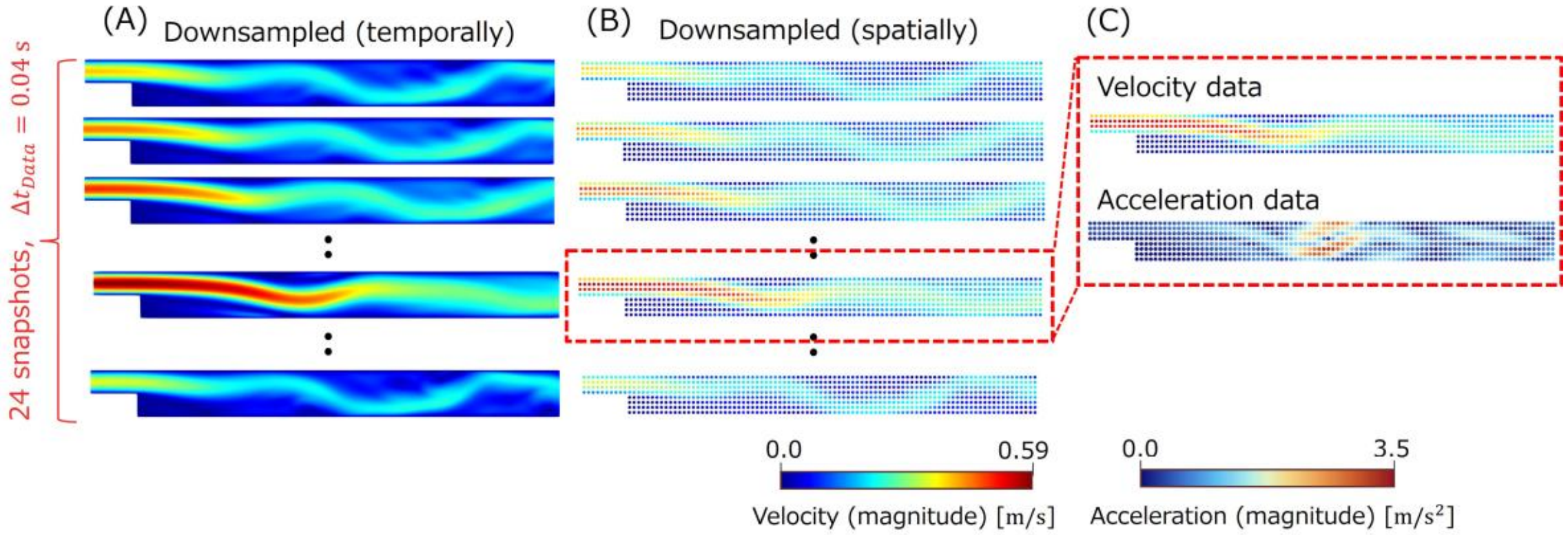

Fig. 4 Generation of synthetic data by spatiotemporal downsampling of the reference velocity fields: (A) temporally downsampled velocity fields, (B) spatially downsampled velocity fields, and (C) velocity and acceleration fields at the target time instant $t = 0.2T$.

Fig. 5 Spatial distribution of training points on which each loss is evaluated.

### 2.7 Error metrics

To quantify reconstruction accuracy, we evaluate the mismatch between the PINN predictions and the reference CFD solution on a set of $M$ evaluation points $\{\mathbf{x}_m\}_{m=1}^{M}$ in the spatial domain, sampled from the reference CFD mesh. We report relative $L_2$ errors for velocity, pressure gradient, and acceleration, defined as

$$e_{ref}^{vel} = \left( \frac{\sum_{m=1}^{M} \left( \mathbf{u}_{ref}(\mathbf{x}_m) - \hat{\mathbf{u}}(\mathbf{x}_m; \theta) \right)^2}{\sum_{m=1}^{M} \mathbf{u}_{ref}(\mathbf{x}_m)^2} \right)^{\frac{1}{2}}, \tag{20}$$

$$e_{ref}^{pre} = \left( \frac{\sum_{m=1}^{M} \left( \nabla p_{ref}(\mathbf{x}_m) - \nabla \hat{p}(\mathbf{x}_m; \theta) \right)^2}{\sum_{m=1}^{M} \nabla p_{ref}(\mathbf{x}_m)^2} \right)^{\frac{1}{2}}, \tag{21}$$

$$e_{ref}^{acc} = \left( \frac{\sum_{m=1}^{M} \left( \mathbf{a}_{ref}(\mathbf{x}_m) - \hat{\mathbf{a}}(\mathbf{x}_m; \theta) \right)^2}{\sum_{m=1}^{M} \mathbf{a}_{ref}(\mathbf{x}_m)^2} \right)^{\frac{1}{2}}. \tag{22}$$

Here, $\mathbf{u}_{ref}(\mathbf{x}_m)$, $\nabla p_{ref}(\mathbf{x}_m)$, and $\mathbf{a}_{ref}(\mathbf{x}_m)$ are the CFD velocity, pressure gradient, and acceleration evaluated at $\mathbf{x}_m$, respectively.

## 3. Results and Discussion

### 3.1 Effect of temporal sampling

In this section, we examine the effect of temporal sampling of velocity snapshots on the reconstruction accuracy of the proposed PINN framework. All loss weights are set to one ($w_{Data}^{vel} = w_{Data}^{acc} = w_{BC} = w_{PDE} = 1$), and three sampling intervals are tested: $\Delta t_{Data} = 0.04, 0.1$, and $0.20$ s.

Figure 6 shows the training histories of the total loss for all $\Delta t_{Data}$. In all cases, the total loss decreases rapidly in the early epochs and then gradually converges.

Figure 7 visualizes the predicted velocity, pressure-gradient, and acceleration fields at the target time ($t = 0.2T$) for different temporal sampling intervals, compared against the CFD reference solutions. For all $\Delta t_{Data}$ considered, the predicted velocity and pressure-gradient fields exhibit good agreement with the reference. In contrast, the acceleration prediction is highly sensitive to temporal sampling interval. As $\Delta t_{Data}$ increases, the predicted acceleration field deviates from the reference solution and becomes increasingly biased toward the acceleration data used during training.

Figure 8 summarizes the relative $L_2$ errors in velocity, pressure gradient, and acceleration ($e_{ref}^{vel}$, $e_{ref}^{pre}$, and $e_{ref}^{acc}$) for different temporal sampling intervals. The velocity reconstruction maintains low $e_{ref}^{vel}$ across all sampling intervals, indicating weak sensitivity to temporal downsampling. Similarly, $e_{ref}^{pre}$ remains at a moderate value and varies only slightly with $\Delta t_{Data}$. These trends indicate that the velocity and pressure gradient reconstructions are only weakly affected by the sparsity of temporal sampling. In contrast, $e_{ref}^{acc}$ increases with $\Delta t_{Data}$.

The increase in $e_{ref}^{acc}$ is expected given how the acceleration data is created. The acceleration data is obtained by differentiating cubic-spline interpolants fitted to temporally downsampled velocity snapshots. As the temporal sampling becomes coarser, interpolation errors increase, limiting the achievable accuracy of the inferred acceleration field even though the acceleration-mismatch loss is intended to promote temporal consistency.

Importantly, this reduced acceleration accuracy does not affect the key hemodynamic quantities. Across all tested $\Delta t_{Data}$, the proposed framework consistently reconstructs the velocity field with low relative $L_2$ error and weak sensitivity to temporal downsampling, which is critical because velocity directly determines clinically and physiologically meaningful metrics such as flow rate and wall shear stress. In addition, the pressure-gradient field also remains robust to temporal downsampling, supporting reliable assessment of driving forces and associated hemodynamic loads. Therefore, while dense temporal sampling is beneficial when accurate acceleration is required, the proposed approach remains effective for recovering the primary hemodynamic fields of interest even from sparsely sampled velocity snapshots.

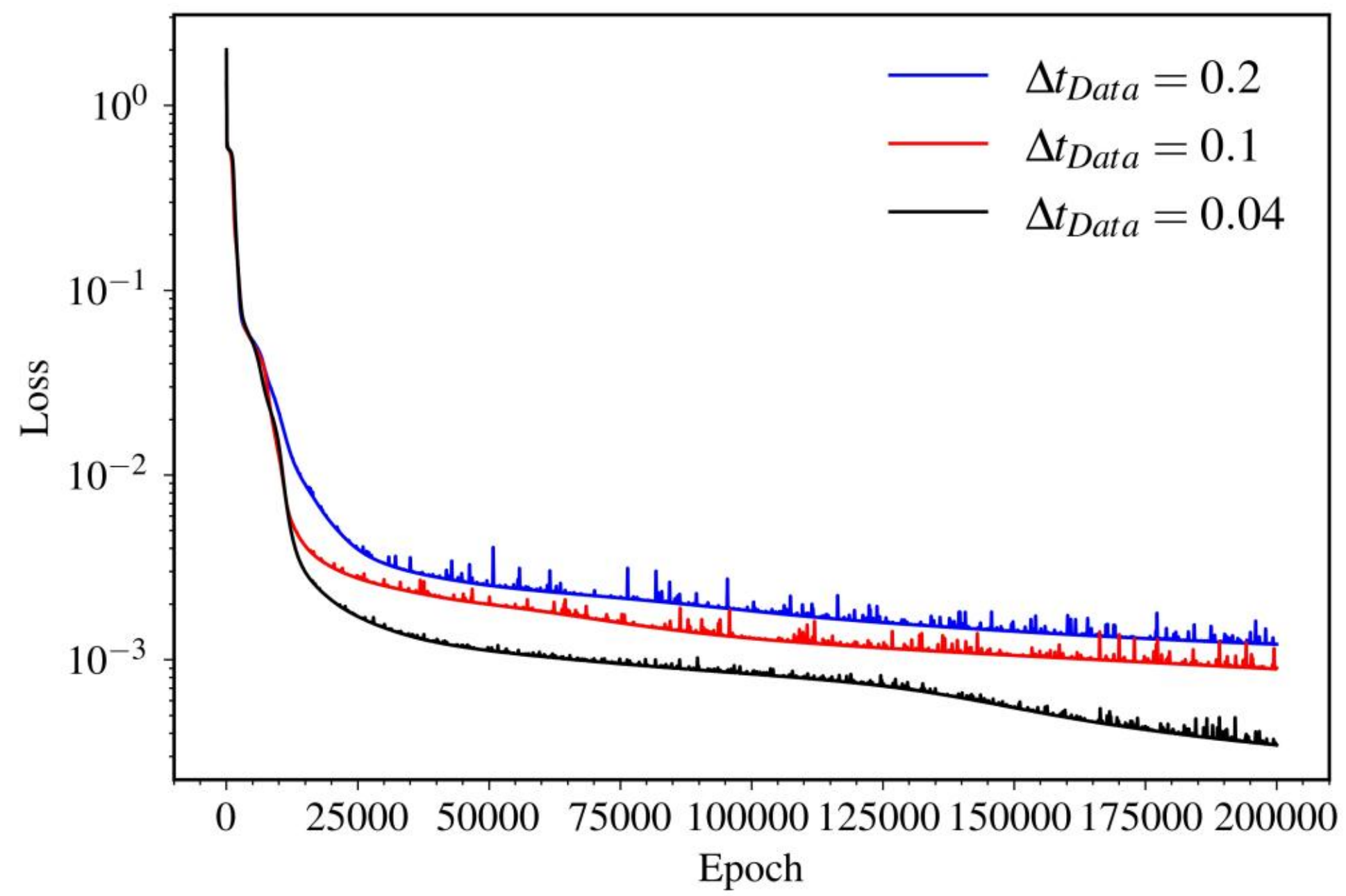

Fig. 6 Training histories of the total loss for different values of temporal sampling intervals ($\Delta t_{Data}$).

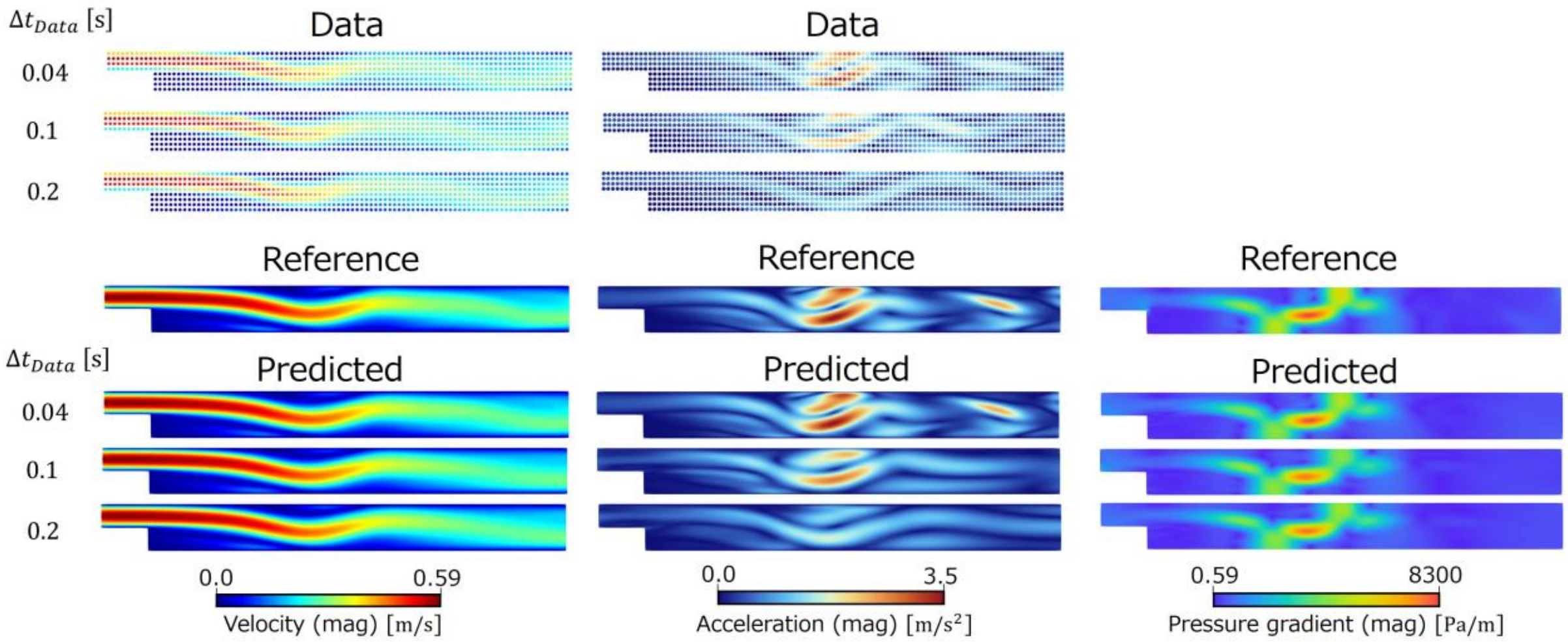

Fig. 7 Predicted velocity, acceleration, and pressure-gradient fields for different values of temporal sampling intervals ($\Delta t_{Data}$).

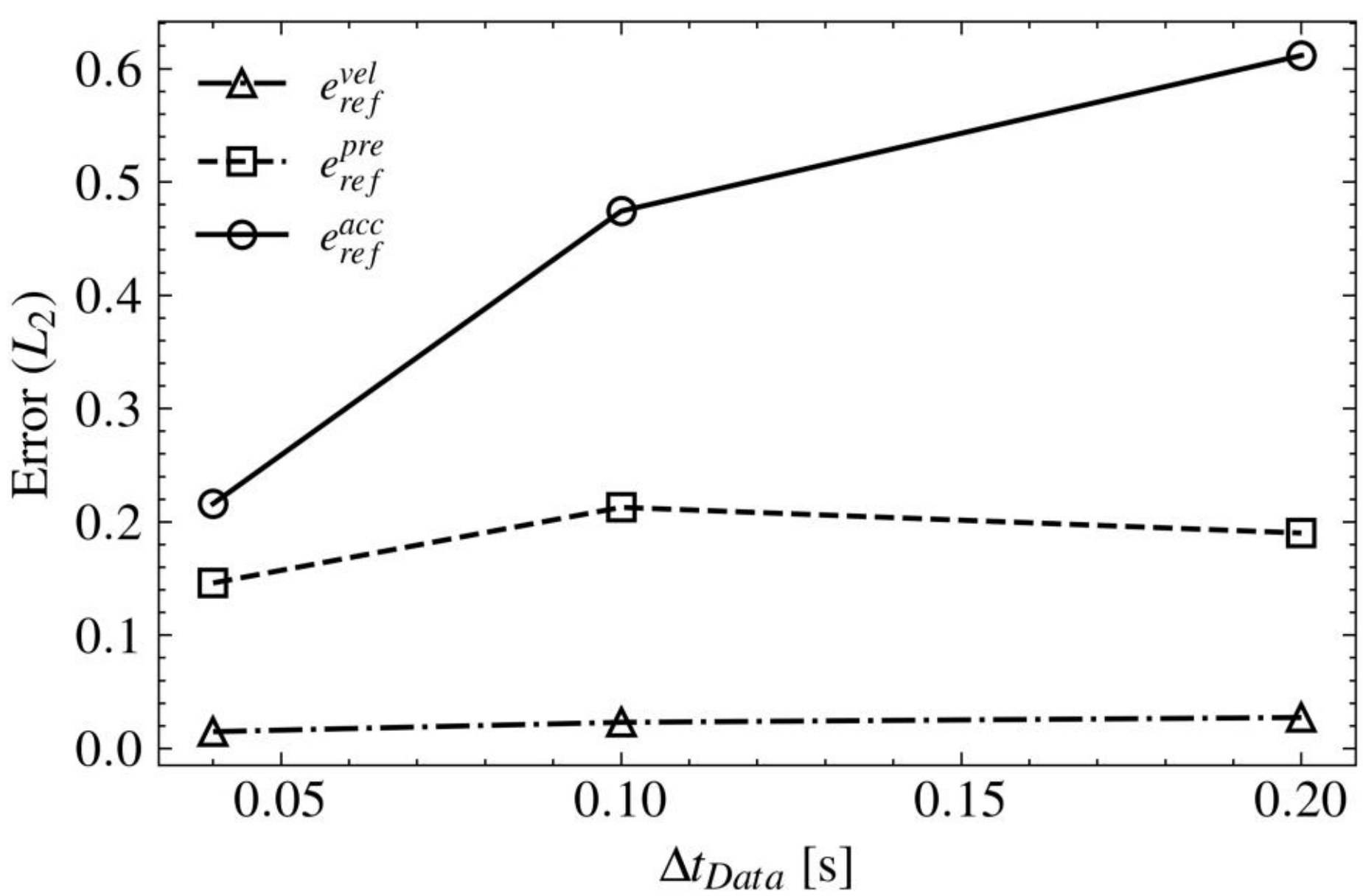

Fig. 8 Relative $L_2$ errors in predicted velocity, acceleration, and pressure-gradient fields for different values of temporal sampling intervals ($\Delta t_{Data}$).

**3.2 Sensitivity to loss weight**

We next examine the sensitivity of the proposed framework to the loss weights, focusing on the acceleration-mismatch weight $w_{Data}^{acc}$. Here, we fix the temporal sampling interval to $\Delta t_{Data} = 0.20\ s$ (the sparsest sampling case in Section 3.1) and vary $w_{Data}^{acc} = 0.0,\ 0.001,\ 0.01,\ 0.1,\ 1.0,$ and $10$, while keeping all other weights equal to one.

Figure 9 compares the predicted velocity, pressure gradient, and acceleration fields for different values of $w_{Data}^{acc}$. The velocity predictions agree well with the reference across all weights, suggesting that the velocity reconstruction is relatively insensitive to $w_{Data}^{acc}$ in this setting. Meanwhile, the acceleration and pressure gradient fields are more sensitive to the choice of $w_{Data}^{acc}$. For larger weights ($w_{Data}^{acc} = 1$ and 10), the predicted acceleration closely follows the provided acceleration data. However, since the data are derived from temporally sparse velocity snapshots, they end up deviating from the reference acceleration field. For a moderate weight $w_{Data}^{acc} = 0.1$, the predicted acceleration field more closely matches the reference. As $w_{Data}^{acc}$ decreases toward zero, the acceleration prediction degrades. The pressure-gradient field shows a similar trend: it remains in good agreement with the reference for moderate weights, but degrades as $w_{Data}^{acc}$ decreases toward zero.

Figure 10 summarizes the relative $L_2$ errors in velocity, pressure gradient, and acceleration ($e_{ref}^{vel},\ e_{ref}^{pre},$ and $e_{ref}^{acc}$) for different values of $w_{Data}^{acc}$. The velocity error $e_{ref}^{vel}$ remains low over the entire range of weights, consistent with the qualitative results in Fig. 8. In contrast, $e_{ref}^{pre}$ and $e_{ref}^{acc}$ vary with $w_{Data}^{acc}$ and become especially large when $w_{Data}^{acc} = 0$. The acceleration error $e_{ref}^{acc}$ is minimized around $w_{Data}^{acc} = 0.1$, and $e_{ref}^{pre}$ is also relatively low for moderate weights.

These results suggest that accurate reconstruction remains feasible even when the acceleration data include interpolation errors, provided that $w_{Data}^{acc}$ is tuned appropriately. This requirement is particularly practical for 4D-flow MRI, where the temporal resolution is typically $\Delta t_{\mathrm{Data}} \leq 0.1$ s.

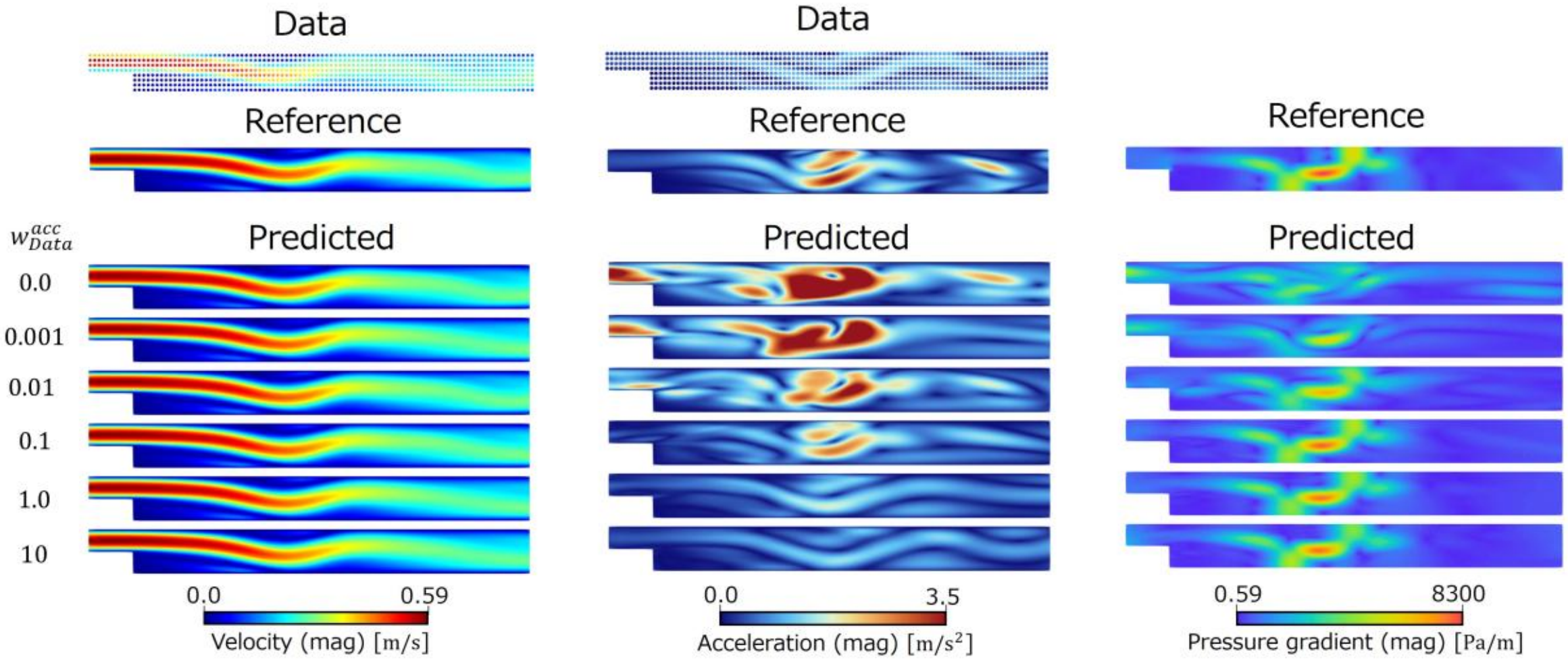

Fig. 9 Predicted velocity, acceleration, and pressure-gradient fields for different values of the acceleration-mismatch weight ($w_{Data}^{acc}$).

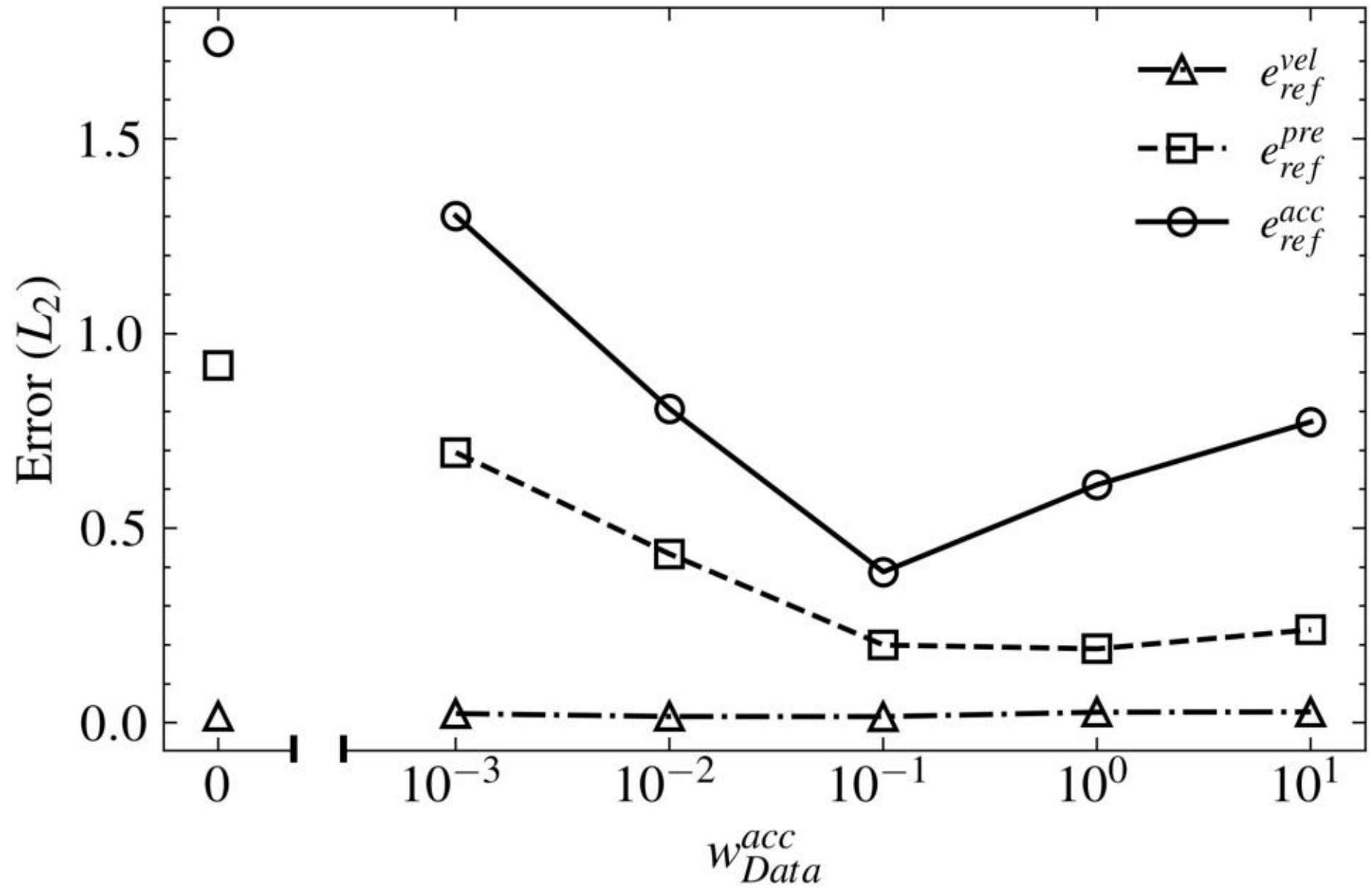

Fig. 10 Relative $L_2$ errors in predicted velocity, acceleration, and pressure-gradient fields for different values of the acceleration-mismatch weight ($w_{Data}^{acc}$). The horizontal axis is logarithmic for $w_{Data}^{acc} > 0$ and the case $w_{Data}^{acc} = 0$ is shown as a baseline.

### 3.3 Limitations and future considerations

Finally, we summarize limitations of the present study and possible extensions. In this study, we considered a two-dimensional flow with a rigid wall. Extension to fully three-dimensional cases and flows with wall motion remains to be investigated. Moreover, the proposed approach has not been evaluated for blood flows with higher Reynolds and Womersley numbers (e.g., aortic flows), where the reconstructions may become more challenging. Another practical limitation concerns the quality of the available velocity data: the robustness observed here may not hold when the velocity measurements include noise or when the number of data points is too small, which is often the case with actual MRI data. Future work should therefore focus on applications to MRI measurements and on incorporating additional regularization or measurement-aware loss formulations. For example, the velocity data loss can be defined in terms of voxel-averaged predictions to account for the spatial averaging inherent to MRI acquisition (Fathi et al., 2020). Finally, the current approach reconstructs the flow at a single target time and thus cannot directly provide time-averaged quantities such as time-averaged wall shear stress (Ichimura et al., 2025). A straightforward extension is to train independent models for multiple target instants in parallel and then improve temporal coherence by coupling neighboring phases. In addition, sequential reconstruction may benefit from transfer learning (fine-tuning), where the model trained at one time instant is used to initialize training at the next time instant (Takao and Ii, 2026).

## 4. Conclusion

In this study, we proposed a PINN framework for reconstructing an instantaneous flow field from temporally sparse velocity snapshots by treating the acceleration term in the incompressible Navier-Stokes equations as an additional unknown output of the network. By removing explicit time from the network inputs, the governing equations can be enforced using spatial derivatives alone, which reduces the cost associated with space-time training while retaining physical consistency with transient flow dynamics. In numerical experiments on pulsatile flow behind a stenosis, the proposed framework consistently reconstructed the velocity field with low relative $L_2$ error across a range of temporal sampling intervals. While the acceleration prediction was sensitive to temporal downsampling due to interpolation errors in the computed acceleration data, the velocity and pressure gradient fields remained robust to temporal downsampling. In addition, the sensitivity analysis on the acceleration-mismatch loss weight demonstrated that moderate acceleration regularization improves pressure-gradient and acceleration predictions, even when the acceleration data used in the mismatch term is affected by interpolation errors.

From the standpoint of hemodynamics, the proposed formulation is particularly useful in practical settings where the primary clinical interest lies in quantities at selected phases of the cardiac cycle, such as the peak flow rate (Lala et al., 2025). In such settings, the present approach offers a low-cost and straightforward framework for patient-specific hemodynamic analysis.

## Acknowledgements

This work was supported by JST SPRING, Japan (Grant Number JPMJSP2180), the MEXT Program for Promoting Researches on the Supercomputer Fugaku (Development of human digital twins for cerebral circulation using Fugaku, JPMXP1020230118) and JSPS KAKENHI Grant Number JP22H00190, JP24K02557. Some related preliminary applications were developed using supercomputer Fugaku provided by the RIKEN Center for Computational Science (project ID: hp230208, hp240220, hp250236).